\documentclass[twocolumn,secnumarabic,amssymb, nobibnotes, aps, prd]{revtex4}
\usepackage{graphicx}

\begin{document}
\title{Experimental investigations of the problem of the quantum jump with the help of superconductor nanostructures}
\author{V. L. Gurtovoi}  \author{A. I. Il'in} \author{A.V. Nikulov}
\affiliation{Institute of Microelectronics Technology and High Purity Materials, Russian Academy of Sciences, 142432 Chernogolovka, Moscow District, RUSSIA.} 

\begin{abstract} The quantum theory, that we now know, arose in the process of sharp disputes between its creators. One of the results of these disputes was the emergence in recent years of fundamentally new areas of investigation and technology - quantum information and quantum computing. One of the subjects of controversy between the creators of quantum theory was the quantum jump. We draw the attention to the possibility of experimental investigation of this problem with the help of quantum superconductor nanostructures. The first results of experiments are presented, the paradoxicality of which indicates the relevance of the problem. 
 \end{abstract}

\maketitle

\narrowtext

\section{Introduction}
\label{}
Quantum mechanics is considered fairly to be the most successful theory of physics. But its history is extremely dramatic. Quantum mechanics arose as a result of fierce disputes between its creators, Planck, Einstein, Schrodinger, and others on the one hand, and Bohr, Heisenberg, Dirac, and others on the other hand. This dispute remains relevant to this day. The title "Closing the Door on Einstein and Bohr's Quantum Debate" of the Viewpoint of Alain Aspect published recently \cite{Aspect2015} does not mean that the quantum debate is really over. The debate continues. The quantum debate between Einstein and Bohr, which became particularly relevant thanks to experimental evidences of violation \cite{BellTest1982,BellTest2015,BellTest2017,BellTest2018} of Bell's inequalities \cite{Bell1964}, resulted to the emergence of new important areas of research and technology - quantum information and quantum computing \cite{QuantComp2000}. The ideas of quantum information and quantum computing are based on the EPR (Einstein - Podolsky Rosen) correlation, the most paradoxical quantum principle which emerged due to the debate about the question "Can Quantum-Mechanical Description of Physical Reality be Considered Complete?" \cite{EPR1935,Bohr1935}.  

This question was not the only one in the disputes between the creators of the quantum theory. The controversy about quantum jumps was no less fierce. The quantum jumps are implied in the model of the atom proposed by Bohr in 1913. Bohr postulated that orbits of electron in atom can be stationary because its angular momentum $pr$ should be divisible by the Planck constant $\hbar = h/2\pi $
$$pr = n\hbar \eqno{(1)}$$   
The quantum jumps are implied since the quantum number $n$ can change only on an integer number. Schrodinger hoped to get rid of the quantum jumps using his wave mechanics \cite{Schrodinger1926}. "{\it He objects in particular to the notion of  'stationary states', and above all to 'quantum jumping' between those  states. He regards these concepts as hangovers from the old Bohr quantum  theory, of 1913, and entirely unmotivated by anything in the mathematics of the new theory of 1926. He would like to regard the wavefunction itself as the complete picture, and completely determined by the Schrodinger equation, and so evolving smoothly without 'quantum jumps'}" \cite{Bell1987}. 

Heisenberg recalled \cite{Heisenberg1971} the heated discussions between Bohr and Schrodinger in the same year 1926, when Schrodinger published his wave mechanics. Schrodinger tried to convince Bohr: "{\it Surely you realize that the whole idea of quantum jumps is bound to end in nonsense}" \cite{Heisenberg1971}. He was in despair at the impossibility of getting rid of the jumps: "{\it If all this damned quantum jumping were really here to stay, I should be sorry I ever got involved with quantum theory}" \cite{Heisenberg1971}. After 26 years, Schrodinger tried to explain his negative attitude to quantum jumps in an article "Are there quantum jumps?" \cite{Schrodinger1952}. 

Superconductor nanostructures are used in this work for experimental investigations of the problem of the quantum jump. We remind in the next Section that the Bohr quantization is observed not only in atoms, but also in nanostructures, for example in nano-rings, in order to substantiate the possibility of using nanostructures for experimental investigation of the problem of the quantum jumps. Electrons, being fermions, occupy the levels with different quantum numbers. Therefore it is difficult to use  normal metal nano-rings for experimental investigation of the quantum jumps which should be observed due to a change of quantum number. 

We draw reader's attention in the third section that Cooper pairs, in contrast to electrons, have the same quantum number in a superconducting ring. Moreover,  Cooper pairs  cannot change their quantum number individually. Therefore a superconducting ring is ideal object for experimental investigation of the quantum jumps. Numerous experimental facts considered in the fourth section give evidence that the quantum number $n$ of Cooper pairs changes with the variation of the magnetic flux inside the superconducting ring because of  the Aharonov - Bohm effect. 

The method of the measurement and the theory of the critical current of superconducting rings are described in the fifth section. The results of measurements of asymmetric superconducting rings with different cross-sections of the halves are compare with theoretical prediction in the sixth section. The quantum jumps of the critical current are not observed contrary to the theory. The experimental results considered in the seven section testify that the quantum jumps are not observed at measurements also of  superconducting rings with asymmetric link-up of current leads in spite of the experimental evidence of the change of the quantum number.  The absence of the quantum jumps of the critical current, contrary to the prediction of the quantum theory, is especially strange because of the observations of similar jumps in other cases considered in the eighth section. In the Conclusion we draw reader's attention that first results of measurements testify about the perspectivity to use superconducting structures of different shapes for experimental investigations of the fundamental problem of the quantum jumps.

\section{The Bohr quantization in nanostructures}
\label{}
Now almost no one doubts the existence of the quantum jumps, just as only a few doubt that the quantum-mechanical description is complete. But the experience of the emergence of quantum computer science suggests that attention to the subject of the dispute between the creators of quantum theory can lead to important breakthroughs in science and technology. Now the possibilities of experimental research are incomparably greater than they were in 1926 and even in 1952. We know that phenomena connected with the Bohr quantization (1) is observed not only in atoms, but also in nanostructures. These phenomena are observed when the energy difference 
$$E_{n+1} - E_{n} = \frac{p_{n+1}^{2}}{2m} - \frac{p_{n}^{2}}{2m} = (2n+1)\frac{\hbar ^{2}}{2mr^{2}} \eqno{(2)}$$ 
between adjacent permitted states $n+1$ and $n$ increases the thermal energy $k_{B}T$.  The energy spectrum of atom is strongly discrete due to small radius of electron orbits: the energy difference between adjacent permitted states $\Delta E \approx \hbar ^{2}/2mr^{2} \approx 2 \ 10^{-18} \ J$ for the Bohr radius $r_{B} \approx 0.05 \ nm = 5 \ 10^{-11} \ m$. This energy difference corresponds to the very high temperature $T = \Delta E/k _{B} \approx 100000 \ K$. The temperature decreases by four orders of magnitude, to $T = \Delta E/k _{B} \approx 10 \ K$, when the quantization radius increases to $r \approx 5 \ nm = 5 \ 10^{-9} \ m$ and by eight orders of magnitude, down to $T = \Delta E/k _{B} \approx 0.001 \ K$ at a radius of $r \approx 500 \ nm = 5 \ 10^{-7} \ m$.

The persistent current, phenomenon connected with the Bohr quantization (1) is observed in normal metal nano-rings with a radius $r > 300 \ nm$ at the temperature $T \approx 1 \ K$  \cite{PCScien09,PCPRL09} because electrons at the Fermi level, which create this current, have very great quantum number $n_{F} \gg 1$ and therefore $(2n+1)\hbar ^{2}/2mr^{2}k _{B} \approx 1 \ K$. The persistent current of electrons \cite{PCScien09,PCPRL09} is created by one electron on the Fermi level $n _{F}$ per one - dimensional channel, since electrons, being fermions, occupy the levels from $n = -n_{F}$ to $n = +n_{F}$ with the opposite direction of the velocity (1) \cite{PC1988Ch1}. The persistent current is observed at $T  \approx 1 \ K \gg 0.001 \ K$ in the normal metal rings with $r \approx 500 \ nm $ \cite{PCScien09,PCPRL09} because of the stronger discreteness of the spectrum at the Fermi level $\Delta E _{n_{F}+1,n_{F}} \approx (\hbar^{2}/2mr^{2})2n_{F} \gg \hbar^{2}/2mr^{2}$. The total current of the ring $I _{p,t} \approx \surd N _{ch}I _{p,1}$ scales as $\surd N _{ch}$ rather than as $N _{ch}$ - the number of one - dimensional channels because of random sign of $I _{p,1}$. Free electron, which is not dissipated, should create the current of the order $I _{p,1} \approx ev_{F}/2\pi r \approx 100 \ nA$ in the one - dimensional channel with the $r \approx 300 \ nm $ at a typical value of the Fermi velocity $v_{F} \approx n_{F}\hbar /mr$. The amplitude of the persistent current $I _{p,A} < 1 \ nA$ observed in the normal metal rings with $r \approx 300 \ nm $ \cite{PCScien09} is by several orders of magnitude less than $I _{p,t} \approx \surd N _{ch}100 \ nA$ because of the electron scattering.  Its value decreases exponentially with temperature increasing, see Fig.3 of \cite{PCScien09}. The visible persistent current of electrons may be expected to observe at the room temperature $T \approx 300 \ K $ in rings with a radius  $r \approx 30 \ nm $.   

\section{The persistent current in superconducting rings}
\label{}
This phenomenon can hardly be used to experimental investigation of the problem of quantum jumps since the persistent current in normal metal nano-rings is created by a lot of electrons having different quantum numbers in one-dimensional channels. Superconductors are much better suited for experimental investigation of this problem, since all Cooper pairs in the ring have the same quantum number $n$.

The quantum number $n$ describes the angular momentum of each pair $pr = n\hbar $. Therefore the momentum $p = n\hbar /r$ is microscopic in the ring with the radius $r$. But Cooper pairs in the ring cannot change their quantum number $n$ individually \cite{FPP2009}. This impossibility of individual motion of quantum particle was postulated first by Lev Landau as far back as 1941 \cite{Landau41} for the description of superfluidity of $^{4}He$ liquid in order to explain the observation of macroscopic quantum phenomena: superfluidity and superconductivity. The Landau postulate \cite{Landau41} is used in the Ginzburg-Landau theory \cite{GL1950}. The Ginzburg-Landau wave function $\Psi _{GL} = |\Psi |\exp{i\varphi }$ is similar the Shrodinger wave function. The phase $\varphi $ of the Ginzburg-Landau wave function, as the Shrodinger one, describes the microscopic momentum of each particle $p = \hbar \nabla \varphi $.  But the value $|\Psi |^{2}$  describes the density $n_{s}$ of all Cooper pairs in a region of superconductor rather than the probability to observe a particle in the region. All Cooper pairs in the ring have the same phase $\varphi $ and the same angular momentum according to the Landau postulate \cite{Landau41} and the Ginzburg-Landau theory \cite{GL1950}. 

The angular momentum is determined by the requirement that the complex wave function must be single-valued in any point $l$ of the ring $\Psi = |\Psi |e^{i\varphi } =  |\Psi |e^{i(\varphi + n2\pi )} $, from which the Bohr quantization is deduced 
$$\oint_{l}dl p = \hbar \oint_{l}dl \nabla \varphi  = 2\pi \hbar n \eqno{(3)}$$ 
The differences of the velocity $v = p/m = \hbar n/rm$ between adjacent permitted states $n+1$ and $n$ decrease with the increase of the particle mass $m$. Therefore it is important that the value of the velocity $v = p/m = \hbar n/rm$ is determined by the microscopic mass $m$ of each Cooper pair whereas the energy difference between $n+1$ and $n$ is determined the macroscopic mass $M = mN_{s}$ of superconducting condensate since the quantum number $n$ of Cooper pair can change only simultaneously. Therefore the energy difference 
$$E_{n+1} - E_{n} =\frac{Mv_{n+1}^{2}}{2} - \frac{Mv_{n}^{2}}{2} = N_{s}(2n+1)\frac{\hbar ^{2}}{2mr^{2}} \eqno{(4)}$$
between adjacent permitted states of superconducting ring is more by a multiplier equal to the total number of Cooper pairs $N_{s}$. This number $N_{s}  = \int_{V}dV |\Psi |^{2} = \oint_{l}dl sn_{s} = 2\pi r sn_{s} \gg 1$ is very big in the real ring with a section $s$ and the circumference $l = 2\pi r$. The discreteness (4) increases with the increase of all three sizes of the ring $\Delta E \approx N _{s}\hbar ^{2}/2mr^{2} \approx  n_{s}s2\pi r(\hbar ^{2}/2mr^{2}) \propto  (s/r)$ due to the increase of the number of Cooper pairs $N _{s}$.   

The strong discreteness of the permitted states of superconducting rings and closed loops has been corroborated by numerous experimental results, in particular quantum periodicity in the persistent current of Cooper pairs \cite{LP1962,NANO2002,Letter2003,Science2007,Letter2007,JETP07J,PCJETP07,FNT2010,PL2012PC,APL2016,PLA2017}. According to the canonical definition, the gradient operator $\hat{p} = -i\hbar \nabla $  corresponds to the canonical momentum $p = mv + qA$ of a particle with a mass $m$ and a charge $q$ both with $A \neq 0$ and without $A = 0$ magnetic field and the velocity operator $\hat{v} = (\hat{p} - qA)/m = (-i\hbar \nabla - qA)/m$ \cite{LandauL} depends on the magnetic vector potential $A$. The effects connected with this dependence were first predicted by Aharonov and Bohm \cite{AB1959}. Therefore, they are referred as the Aharonov - Bohm effects. The velocity 
$$\oint dl v = \frac{2\pi \hbar }{m}(n - \frac{\Phi}{\Phi _{0}} ) \eqno{(5)}$$
of a particle with a charge $q$ cannot be equal zero when the magnetic flux $\Phi = \oint_{l}dl A$ inside the ring is not divisible $\Phi \neq n\Phi _{0}$ by the flux quantum $\Phi _{0} = 2\pi \hbar /q$ because of the Bohr quantization and the Aharonov-Bohm effect. The flux quantum is equal $\Phi _{0} \approx 41.4  \ Oe \ \mu m^{2}$ for electrons $q = e$ and $\Phi _{0} \approx 20.7  \ Oe \ \mu m^{2}$ for Cooper pairs $q = 2e$.  

The persistent current observed \cite{PCJETP07} in superconductor   
$$I_{p} =  sqn _{s}v = \frac{qv_{n}N _{s}}{2\pi r} = I_{p,A}2(n - \frac{\Phi }{\Phi_{0}}) \eqno{(6)}$$ 
is much larger than in the normal metal rings \cite{PCScien09,PCPRL09} since all $N _{s}$ Cooper pairs are on the same quantum state $n$. Here $I_{p,A} = \Phi_{0}/2L_{k}$ is the amplitude of the persistent current when the value $(n - \Phi/\Phi_{0})$ changes between -0.5 and +0.5; $L_{k} = ml/sq^{2}n _{s}$ is the kinetic inductance of the ring with the length $l = 2\pi r$, the section $s = wd$ and the density $n _{s} \propto 1 - T/T _{c}$ of Cooper pairs. This macroscopic quantum phenomenon is very suitable for investigating the problem of quantum jumps since the persistent current (6) should change by jump on a macroscopic value when the quantum number $n$ changes by one. The persistent current depends on the quantum number $n$ and the magnetic flux $\Phi = BS + L_{f}I_{p}$ created by the external magnetic field $B$ inside the ring with the area $S = \pi r^{2}$ and by the persistent current $I_{p}$. The kinetic inductance $L _{k} \approx (\lambda _{L}^{2}/s) \mu _{0}l$ exceeds the magnetic inductance $L_{f}  \approx \mu _{0}l$ of a ring with the small cross section $s \ll \lambda _{L}^{2}$ where, $\lambda _{L} = (m/\mu _{0}q^{2}n_{s})^{0.5} = \lambda _{L}(0)(1 - T/T_{c})^{-1/2}$ is the London penetration depth, $\lambda _{L}(0) \approx 50 \ nm = 5 \ 10^{-8} \ m$ for most superconductors \cite{Tinkham}. Such ring with weak screening is more suitable for investigating the problem of quantum jumps. 

\section{Changes of the quantum number of superconducting rings}
\label{}
The energy of the magnetic field induced by the persistent current is less than the kinetic energy of the ring with a small cross-section $s \ll \lambda _{L}^{2}$ \cite{Tinkham}. The total energy of such ring is approximately equal to its kinetic energy \cite{Tinkham}. The kinetic energy of the persistent current, deduced in \cite{PL2012Th}, 
$$E_{n} = \frac{L_{k}I_{p}^{2}}{2} =  I _{p,A}\Phi_{0}(n - \frac{\Phi }{\Phi_{0}}) ^{2} \eqno{(7)}$$  
depends on the value of the persistent current $I_{p}$ and the kinetic inductance $L _{k} \approx (\lambda _{L}^{2}/s) \mu _{0}l$ of the ring. The discreteness of the energy spectrum (7) is determined by the amplitude $I _{p,A}$ of the persistent current (6). The energy difference $\Delta E _{n+1,n} = E _{n+1} - E _{n} \approx I _{p,A}\Phi_{0}$ between two states $n = n'$ and $n = n'+1$ with minimal energy at $n'\Phi_{0} < \Phi < (n'+1) \Phi_{0}$ corresponds to a high temperature $I _{p,A}\Phi_{0}/k _{B}\approx 1500 \ K $  for a typical value of the amplitude $I _{p,A} \approx 10 \ \mu A$ \cite{NanoLet2017}. Numerous experimental results corroborate that the quantum number $n$ takes an overwhelming probability $P \propto exp(-E _{n}/k _{B}T)$ of an integer value corresponding to the minimum kinetic energy at this value of the magnetic flux $\Phi $. 

The quantum number $n$ corresponds to the minimal energy (7) and the maximal probability $P _{n}$ in the interval of the magnetic flux inside the ring $(n'-0.5) < \Phi < (n'+0.5)\Phi_{0}$. The two state $n = n'$ and $n = n'+1$ have the same value of the kinetic energy in (7) $E _{k}= (n\Phi_{0} - \Phi )^{2}/2L_{k} = \Phi_{0}^{2}/8L_{k}$ at $\Phi = (n'+0.5)\Phi_{0}$. According to the universally recognized explanation \cite{Tinkham} the quantum periodicity in the transition temperature \cite{LP1962}, the ring resistance \cite{Letter2007,FNT2010}, the magnetic susceptibility \cite{Science2007}, the dc voltage \cite{NANO2002,Letter2003,PL2012PC,APL2016} and the critical current \cite{JETP07J,PCJETP07,PLA2017} are observed due to the change of the quantum number $n$ with the magnetic flux at $\Phi = (n'+0.5)\Phi_{0}$. 

But the jump with the change of the quantum number from $n = n'$ to $n = n'+1$ should not be observed at the measurement of most of these quantities. The jump should not be observed at measurement of the fractional depression of the transition temperature $\Delta T _{c}/T _{c} \propto  -E _{k} \propto -(n - \Phi/\Phi_{0})^{2}$ \cite{LP1962} and the resistance $\Delta R(\Phi) \propto (n - \Phi /\Phi_{0})^{2}$ \cite{Letter2007,FNT2010} since the variation of these quantities is proportional to $(n - \Phi /\Phi_{0})^{2}$ \cite{Tinkham}. The jump of the magnetic susceptibility measured in the fluctuation region \cite{Science2007} and the dc voltage \cite{NANO2002,Letter2003,PL2012PC,APL2016} should not be observed since these quantities is proportional to the average value $\overline{n - \Phi /\Phi_{0}}$: $n - \Phi /\Phi_{0} = -0.5$ at $n = n'$ and  $n - \Phi /\Phi_{0} = +0.5$ at $n = n'+1$ when $\Phi = (n'+0.5)\Phi_{0}$, but $\overline{n - \Phi /\Phi_{0}} \approx P(n'?)(-0.5) + P(n'?+1)(+0.5) = 0$ since $P(n') = P(n'+1)$ when $\Phi = (n'+0.5)\Phi_{0}$.

\section{The critical current of superconducting rings}
\label{} 
We used measurements of the magnetic dependence of the critical current of superconducting rings of different shapes and ring structures in order to investigate experimentally the problem of quantum jumps. The  experimental setup in all cases was the same. The critical currents in the positive $I_{c+}$ and negative $I_{c-}$ directions with respect to the external measuring current $I_{ext}$ were determined \cite{JETP07J,PCJETP07} by measuring periodically repeating current-voltage characteristics (a period of 0.1 s) in a slowly varying magnetic field $B$ (a period of approximately 100 s) as follows. First, the condition that the structure was in the superconducting state was checked. Next, after the threshold voltage was exceeded (this voltage, set above induced voltages and noises of the measuring system, determined the measured critical current), the values of the magnetic field $B$ and the critical current $I_{c+}$ (or $I_{c-}$) were recorded. The external magnetic field $B$ was induced by a solenoid.   

\begin{figure}
\includegraphics{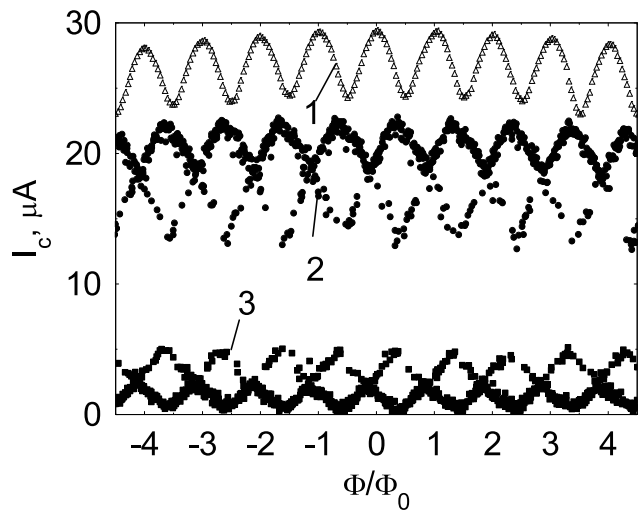}
\caption{\label{fig:epsart} The measurements of the critical current testify most obviously that the quantum number $n$ takes the integer value corresponding to the minimum kinetic energy (7) with  the overwhelming probability. The curve 1 shows the typical magnetic dependence of the critical current $I_{c}(\Phi /\Phi _{0})$ measured on the symmetric aluminum ring. Each triangle corresponds one record of $I_{c+}$ and $B$. The quantum number $n$ takes the single integer number $n$ at a given magnetic flux $\Phi /\Phi _{0}$. The pereodicity of the dependence $I_{c}(\Phi /\Phi _{0})$ testifies that the quantum number changes at $\Phi  = (n'+0.5)\Phi _{0}$. The period of the magnetic field $B_{0} = \Phi _{0}/S \approx 2.1 \ Oe$ corresponds to the area $S = 2\pi ^{2} \approx 10  \ \mu m^{2}$ of the measured ring with the radius $r \approx 1.8 \ \mu m$. The curve 2 shows the critical current measured on the aluminum ring with a small narrow spot. The two different values of the critical current observed at the same magnetic flux $\Phi /\Phi _{0}$ correspond to the two permitted states with the quantum number $n$ and $n+1$. The curve 3 shows  the dependence of the absolute value of the persistent current $|I_{p}| = (I_{c,0} - I_{c})/2$ calculated from the  experimental dependence 2 for the critical current and the theoretical expression (10) when $I_{c,0} = 23 \ \mu A$.}
\end{figure} 

 Each record of $B$ and $I_{c+}$ corresponds to a single point, for example on the curve 2 of Fig.1. The measurements testify that the critical current depends periodically on the external magnetic field $B$ with the period $B_{0} = \Phi _{0}/S$ corresponding to the flux quantum $\Phi _{0} \approx 20.7  \ Oe \ \mu m^{2}$ for Cooper pairs $q = 2e$ inside the ring with the area $S = \pi r^{2}$. In order to emphasize that the oscillation period is equal to the flux quantum we plotted the critical current $I_{c+}$ as the function of the magnetic flux $\Phi /\Phi_{0}$ on Fig.1, Fig.2, Fig.3. The periodical dependencies  leave no doubt that the quantum number changes with the variation of the magnetic flux. The quantum number can change at each measurement of the the critical current. The ring (ring structure) was switched from the superconducting state to the normal state at each $I_{c+}$ (or $I_{c-}$) measurement by the measuring current $I_{ext}$ varying periodically between $I_{ext} < - I_{c-}(B)$ and $I_{ext} > +I_{c+}(B)$. 

The current-voltage characteristics (CVC) of aluminium ring structure used in our work exhibit hysteresis. The typical CVC of these structures is shown on Fig.1 in the previous publication \cite{PRB2014}. Sharp transition of the entire structure from the superconducting state with the zero resistance $R = 0$ to the normal state with the resistance $R > 0$ are observed at $ I_{ext} = I_{c+}$ and $ I_{ext} = -I_{c-}$. The ring structure returns to the superconducting state when the external current decreases to the value $ I_{ext} = I_{rs}$ or $ I_{ext} = -I_{rs}$ which is less than the critical current $ I_{rs} < I_{c+}; I_{c-}$, see Fig.1 in the paper \cite{PRB2014}. The quantum number $n$ in the expressions for the velocity (5) of Cooper pairs and for the persistent current (6) can take any integer value after the ring returns to the superconducting state. The value of the persistent current (6) and, as a consequence of the critical current (10), depend strongly on the quantum number $n$. Therefore the measurements of the critical current testify most obviously that the quantum number $n$ takes the integer value corresponding to the minimum kinetic energy (7) almost in all cases. 

The critical current of a symmetric superconducting ring corresponds to the single integer number $n$ at a given magnetic flux  $\Phi $: $n = - 2$ at $-2.5 \Phi _{0} < \Phi < -1.5 \Phi _{0}$; $n = - 1$ at $-1.5 \Phi _{0} < \Phi < -0.5 \Phi _{0}$; $n = 0$ at $-0.5 \Phi _{0} < \Phi < 0.5 \Phi _{0}$; $n = 1$ at $0.5 \Phi _{0} < \Phi < 1.5 \Phi _{0}$; $n = 2$ at $1.5 \Phi _{0} < \Phi < 2.5 \Phi _{0}$, see, Fig.1, Fig.2. The value of the critical current corresponds more than one integer number at the same magnetic flux $\Phi $ only in rare cases. The experimental dependence 2 on Fig.1, recurring rhombuses, is similar to the theoretical prediction for a symmetric superconducting ring (10) when quantum number $n$ takes different integer numbers $n$ and $n+1$ at the same magnetic flux $\Phi $. 

According to the theoretical prediction (10) the dependence of the the critical current on the magnetic flux $\Phi = n'\Phi _{0} + \delta \Phi $ should be $I_{c} = I_{c0} - I_{p,A}|\delta \Phi /\Phi _{0}|$ when $n = n'$ and $I_{c} = I_{c0} - I_{p,A}(1 - |\delta \Phi /\Phi _{0}|)$ when $n = n'+1$ at $\delta \Phi > 0$ and $n = n'-1$ at $\delta \Phi < 0$. The experimental dependence 2 on Fig.1 testifies that the superconducting states $n = n'$ corresponding to  the minimum kinetic energy (7) have higher probability (which is proportional to the number of the points) but the probability of the other states $n = n'+1$ and $n = n'-1$ is not zero in contrast to the experimental dependence 1 on Fig.1. 

The similarity of the experimental dependence $I_{c}(\Phi /\Phi _{0})$ of the critical current 2 with the theoretical prediction (10) allows to plote the magnetic dependence of the persistent current $|I_{p}| = (I_{c,0} - I_{c})/2$, see the curve 3 on Fig.1. The $|I_{p}|(\Phi /\Phi _{0})$ dependence 3 should be expected according to the experimental data 2 and the expression (10) for the critical current. The experimental dependence 3 differs from the theoretical dependence (6) $|I_{p}| = I_{p,A}2|n - \Phi /\Phi _{0}|$ only by an inexplicable shift on $0.3 \Phi _{0}$, which is observed when the critical current of asymmetric superconducting rings are measured \cite{JETP07J}.

The critical current corresponds to the value of the external current at which the velocity of Cooper pairs reaches a critical value equal to the depairing velocity $v_{sc} = \hbar /m \surd 3\xi (T)$  \cite{Tinkham}. The velocity of Cooper pairs in the segments of the ring depends on the velocity quantization (5), the external current $I_{ext}$, the form of the ring and the geometry of link-up of the current leads. According to the quantization condition (5)
$$l_{u}v_{u} - l_{l}v_{l} = \frac{2\pi \hbar }{m}(n - \frac{\Phi }{\Phi _{0}}) \eqno{(8)} $$
where $l_{u}$ and $l_{l}$ are the length the upper segment and the lower segment of the ring between the current leads, see Fig.2 and Fig.3; $v_{u}$ and $v_{l}$ are the velocity of Cooper pairs in the upper and lower segments of the ring. The external current should be equal the sum
$$I_{ext} = I_{u} + I_{l} = s_{u}j_{u} + s_{l}j_{l} = 2en_{s}(s_{u}v_{u} + s_{l}v_{l}) \eqno{(9)} $$
of the superconducting currents in the upper segment $I_{u}$ and the lower segment $I_{l}$. Here $s_{u}$ and $s_{l}$ are the cross section of the upper and lower segments; $j_{u}$ and $j_{l}$ are the density of superconducting current in the upper and lower segments; $n_{s}$ is the density of Cooper pairs in the ring; $2e$ is the charge of Cooper pair. The direction of the velocity $v_{u}$ and $v_{l}$ and of the external current $I_{ext}$ from left to right and the persistent current (6) clockwise are taken as a positive direction.

The velocities $v_{u}$ and $v_{l}$ are determined by the ratio of segment lengths $v_{u}/v_{l} = l_{l}/l_{u}$ according to (8) and the external current (9) when the persistent current (6) is zero at $n - \Phi /\Phi _{0} = 0$: $v_{u} = I_{ext}l_{l}/2en_{s}(s_{u}l_{l} + s_{l}l_{u})$ and $v_{l} = I_{ext}l_{u}/2en_{s}(s_{u}l_{l} + s_{l}l_{u})$. The velocities equal $v_{u} = v_{l} = I_{ext}/2en_{s}2s_{u}$ in the symmetric ring $s_{u} = s_{l}$ with the symmetric link-up of the current leads $l_{u} = l_{u}$. The external current exceeds the critical value $I_{ext} = I_{c0}$ at $v_{u} = v_{l} =  v_{c}$ and therefore the critical current $I_{c0} = 2en_{s}2s_{u}v_{c}$ when $I_{p} \propto n - \Phi /\Phi _{0} = 0$. The circular persistent current increases the velocity in one of the segment and thus decreases of the critical current. The critical current of the symmetric ring with the symmetric link-up of the current leads should be equal
$$I _{c} = I _{c0} - 2|I _{p}| = I _{c0} - 2I _{p,A}2|n - \frac{\Phi}{\Phi _{0}}| \eqno{(10)}$$
according to (6) and (8). The relative amplitude of the critical current oscillation in a magnetic field increases with decreasing radius of the ring since $I _{p,A}/I _{c0} =  \surd 3\xi (T)/4r$ \cite{PCJETP07}. Here $\xi (T) = \xi (0)(1 - T/T _{c})^{-1/2}$ is the correlation length of the superconductor. The numerous measurements of symmetric rings have corroborated the theoretical prediction (10) when the quantum number $n$ changes by 1 at $\Phi = (n'+0.5)\Phi_{0}$, see Fig.2. Magnetic field dependence of the critical current of the symmetrical ring has the maximums at $\Phi = n'\Phi_{0}$ and the minimums at $\Phi = (n'+0.5)\Phi_{0}$ in accordance with (10) when $n = n'$ in the interval of the magnetic flux $(n'-0.5)\Phi_{0} < \Phi < (n'+0.5)\Phi_{0}$.  

The quantum number takes the value corresponding to the minimal kinetic energy (7), when the ring returns in the superconducting state at $|I_{ext}| < I_{rs}$: $n = n'$ at $\Phi < (n'+0.5)\Phi_{0}$ and  $n = n'? + 1$ at $\Phi > (n'+0.5)\Phi_{0}$. The jump of the persistent current from $I_{p} = -I_{p,A}$ to $I_{p} = +I_{p,A}$ should take place because of this change of the quantum number according to (6). But no jump of the critical current of the symmetric ring should be observed according to (10) since $|-I_{p,A}| = |+I_{p,A}|$. The jump of the critical current can be observed only if an asymmetric ring is measured. The two types of simplest asymmetry are the different cross section of the segments $s_{u} > s_{l}$, Fig.2 and the different length of the segments $l_{u} > l_{l}$, Fig.3. 

\section{Asymmetric superconducting rings}
\label{}
The velocities $v_{u}$ and $v_{l}$ induced only by the external current $I_{ext}$ at $I_{p} \propto n - \Phi /\Phi _{0} = 0$ are equal in the asymmetric ring with $l_{u} = l_{l}$ according to (8). But the currents in the ring halves should be different $|I_{u}| = 2en_{s}s_{u}|v_{u}| > |I_{l}| = 2en_{s}s_{l}|v_{l}|$ in the asymmetric ring with $s_{u} > s_{l}$ when $I_{ext} \neq 0$ and $I_{p} = 0$. In contrast to the external current, the circular persistent current must have the same value $I_{p}$ whereas the velocity of the pairs induced by it must be different $|v_{u}| = |I_{p}|/2en_{s}s_{u} < |v_{l}| = |I_{p}|/2en_{s}s_{u}$ in the halves of the asymmetric ring with different cross sections $s_{u} > s_{l}$. The velocities in the ring halves $v_{u}$ and $v_{l}$ induced by $I_{ext}$ and $I_{p}$ add or subtract depending on the direction of $I_{ext}$ and $I_{p}$ in each of the halves. Therefore the maximum value of the velocity, which determines the value of the critical current, should jump because of the change of the direction inversion of the persistent current with the quantum number change from $n = n'$ to $n = n'? + 1$ at $\Phi = (n'+0.5)\Phi_{0}$.  

According to (8) and (9) the velocities should be equal: in the narrow ring half 
$$v_{l} = \frac{I_{ext}}{2en_{s}(s_{l} + s_{u})} - \frac{2\hbar }{mr}\frac{s_{u}}{s_{l}+s_{u}}(n - \frac{\Phi }{\Phi _{0}}) \eqno{(11a)}$$
and in the wide ring half,
$$v_{u} = \frac{I_{ext}}{2en_{s}(s_{l} + s_{u})} + \frac{2\hbar }{mr}\frac{s_{l}}{s_{l}+s_{u}}(n - \frac{\Phi }{\Phi _{0}}) \eqno{(11b)}$$
when the positive values correspond to the $I_{ext}$ direction from left to right and the $I_{p}$ direction clockwise, Fig.2. The external current reaches the critical value at which the ring transits to the normal normal when the pair velocity in one of the halves reaches the depairing velocity $v_{c} $  \cite{Tinkham}: $|v_{u}| =  v_{c}$ when the external current and the persistent current have the same direction in the upper ring half  and $|v_{l}| =  v_{c}$ when the directions are the same in the lower ring half. 

\begin{figure}
\includegraphics{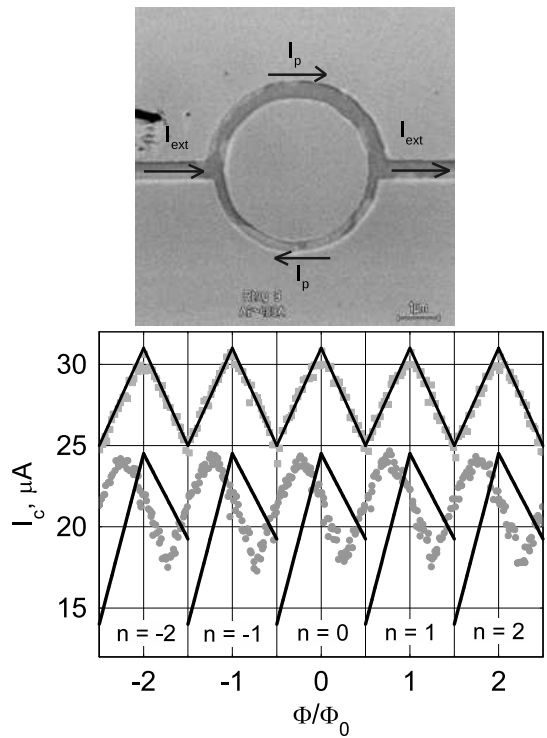}
\caption{\label{fig:epsart}1) The upper picture: Photo of the asymmetric aluminum ring with the radius $r \approx 2 \ \mu m$ and the different width of the halves $w_{w} \approx 0.4 \ \mu m$ and  $w_{n} \approx 0.2 \ \mu m$ is shown. The circular persistent current $I_{p}$ increases the total current in the ring half in which its direction coincides with the direction of the external current $I_{ext}$. The lower picture: The periodic magnetic dependence of the critical current. At the top: The experimental dependence (squares) obtained at the measurement of the symmetric ring with $r \approx 2 \ \mu m$ and the equal width of the halves $w_{w} =  w_{n}  \approx 0.4 \ \mu m$ is described very well by the theory. At the bottom: The qualitative discrepancy is observed between the theoretical predictions (lines) and the experimental dependence (circles) measured on the asymmetric ring shown in the upper figure. The period of the magnetic field $B_{0} = \Phi _{0}/S \approx 1.5 \ Oe$ corresponds to the area $S = 2\pi ^{2} \approx 14 \ \mu m^{2}$ of the measured ring with the radius $r \approx 2.1 \ \mu m$.}
\end{figure} 

Thus, the value of the critical current of the asymmetric ring $s_{u} > s_{l}$ should depend on the direction of both the external current and the persistent current: the critical current $I_{c+}$ measured in the positive direction should be higher than the one $I_{c-}$ measured in the positive direction when the persistent current flows clockwise in the case shown of Fig.2. According to (11) 
$$I_{c+}, I_{c-} = I_{c0} - 2I_{p,A}|n - \frac{\Phi }{\Phi _{0}}|(1 + \frac{s_{l}}{s_{u}}) \eqno{(12a)}$$
when $|v_{l} | = v_{sc}$ and
$$I_{c+}, I_{c-} = I_{c0} - 2I_{p,A}|n - \frac{\Phi }{\Phi _{0}}|(1 + \frac{s_{u}}{s_{l}}) \eqno{(12b)}$$ 
when $|v_{u} | = v_{sc}$. 

The critical current anisotropy predicted be the theory was observed experimentally \cite{PCJETP07}. Moreover the rectification of the ac current was observed \cite{Letter2003} because of this anisotropy \cite{PCJETP07} which varies periodical with magnetic flux inside the ring $\Phi $. But the cause of the anisotropy revealed by the measurements \cite{PCJETP07} is fundamentally different from the cause predicted by the theory (12). According to the theory (12) the anysotropy of the critical current 
$$I_{c,an} =  I_{c+} - I_{c-} = I_{p}(\frac{s_{u}}{s_{l}} - \frac{s_{l}}{s_{u}})  \eqno{(12c)}$$
should change with the magnetic flux $\Phi $ like the persistent current (6). According to the theoretical prediction (12c) the jump of the critical current should be observed if the persistent current jumps at $\Phi = (n'+0.5)\Phi_{0}$ because of the change of the quantum number from $n = n'$ to $n = n'? + 1$.

The predicted anisotropy (12c) can explain the periodical magnetic dependence of the rectified (dc) voltage $V_{dc}$ observed at measurements of asymmetric superconducting rings \cite{Letter2003}. The rectified voltage $V_{dc}$ should be proportional to the average value of the critical current anisotropy $V_{dc}  \propto \overline{I_{c,an}}$ and thus the average value of the persistent current $V_{dc}  \propto \overline{I_{p}}$ according to the theoretical prediction (12c). The average value $\overline{I_{p}}$ should be equal zero not only at $\Phi = n'\Phi_{0}$ where $I_{p} = 0$ at $n = n'$ according to (6) but also at $\Phi = (n'+0.5)\Phi_{0}$ where the state with $I_{p} = 0$ is forbidden but the two permitted states $n = n'$ and $n = n'? + 1$ have opposite directed persistent current (6) and the equal energy (7). All measurements of the rectified voltage $V_{dc}$ of the asymmetric rings \cite{NANO2002,Letter2003,PL2012PC,APL2016} corroborate that the dc voltage equals zero $V_{dc} = 0$ and changes the sign at $\Phi = n'\Phi_{0}$ and $\Phi = (n'+0.5)\Phi_{0}$ in accordance with the theoretical prediction (12c) and $V_{dc}  \propto \overline{I_{c,an}}$. 

The jump because of the quantum number $n$ change should not be observed when quantities, such as the rectified voltage $V_{dc}$ are measured, which are determined by the average value of the persistent current. But the jump of the critical current of the asymmetric ring (12), which corresponds to the single-shot readout of the persistent current, has to be observed. The ring half in which the velocity (11) reaches the critical value changes when the direction of the persistent current (6) inverts at the change of the quantum number from $n = n'$ to $n = n'? + 1$. Therefore the value of the critical current  $I_{c+}$ from (12a) to (12b) and the $I_{c-}$ value from (12b) to (12a) should change and the jump  $\Delta I_{c+} = I_{p,A}(s_{u}/s_{l} - s_{l}/s_{u})$ should be observed at $\Phi = (n'+0.5)\Phi_{0}$, Fig.2. But the measurements \cite{PCJETP07} have revealed that Nature has preferred to avoid jumps. The anisotropy of the critical current is provided by the shift of the magnetic dependence $I_{c+}(\Phi /\Phi_{0})$ and $I_{c-}(\Phi /\Phi_{0})$ on the quarter of the flux quantum: $I_{c,an}(\Phi /\Phi_{0}) = I_{c+}(\Phi /\Phi_{0}) - I_{c+}(\Phi /\Phi_{0}) \approx I_{c}(\Phi /\Phi_{0}+ 0.25) - I_{c}(\Phi /\Phi_{0}-0.25)$ according to the results of measurements \cite{JETP07J,PCJETP07} where $I_{c}(\Phi /\Phi_{0})$ is the magnetic dependence of the critical current of the symmetric ring described by (10). 

\section{Superconducting rings with asymmetric link-up of current leads }
\label{}
The absence of the jumps on the experimental dependence, shown in Fig.2, might have gladdened Schrodinger. In any case, this experimental result makes the investigation of the problem of the jumps in superconductor nanostructues relevant. The absence of the jumps of the critical current not only the rings with different cross-sections of the halves $s_{u} > s_{l}$, but also the rings with asymmetric link-up of current leads $l_{u} > l_{l}$ makes the problem even more actual. The pair velocity induced by the external current $I_{ext}$ at $I_{p} \propto n - \Phi /\Phi _{0} = 0$ in the short segment exceeds the one in the long segment $v_{l} = v_{u}l_{u}/l_{l} > v_{u}$ according to the condition of quantization  (8). The pair velocity reaches the critical value first in the short segment $|v_{sh}| = v_{c} > v_{long}$ at any value $|n - \Phi /\Phi _{0}| \leq 0.5$ in the ring with a strong asymmetry which is determined by the inequality $(l_{u} - l_{l})l_{u}/l^{2} \geq  I_{p,A}/I_{c0}$. The jump of the critical current should be maximal in this case. 

\begin{figure}
\includegraphics{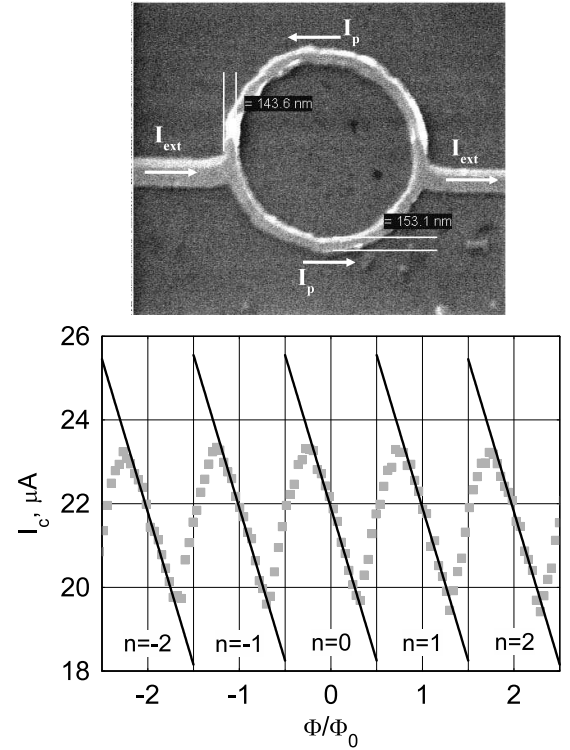}
\caption{\label{fig:epsart} The upper picture: Photo of the aluminum ring with with asymmetric link-up of current leads, the radius $r \approx 1 \ \mu m$, the width $w \approx 150 \ nm$, the length of the longer segment $l_{long} \approx 1.18\pi r$, the length of the short segment $l_{sh} \approx 0.82\pi r$. The circular persistent current $I_{p}$ decreases the critical current when its direction coincides with the direction of the external current $I_{ext}$ in the short segment $l_{sh}$ and increases when the directions $I_{p}$ and $I_{ext}$ are opposite in this segment. The lower picture: The comparison of the theoretical predictions (lines) with the experimental dependence (squares) measured at the temperature $T \approx 1.284 \ K \approx 0.88T_{c}$ on the ring shown in the upper figure. The theoretical dependence corresponds to the critical current $I_{c0} \approx 22 \ \mu A$ at $I_{p} = 0$ and the amplitude of the persistent current  $I_{p,A} \approx 2.2 \ \mu A$. The period of the magnetic field $B_{0} = \Phi _{0}/S \approx 5.7 \ Oe$ corresponds to the area $S = 2\pi ^{2} \approx 3.6 \ \mu m^{2}$ of the measured ring with the radius $r \approx 1.1 \ \mu m$.}
\end{figure}

In the common case the velocity of Cooper pairs should be equal 
$$v_{l} = \frac{l_{u}}{l}\frac{I_{ext}}{qsn_{s}} - \frac{2\pi \hbar }{ml}(n -  \frac{\Phi }{\Phi _{0}}) \eqno{(13a)}$$
in the short segment and 
$$v_{u} =  \frac{l_{l}}{l}\frac{I_{ext}}{qsn_{s}} + \frac{2\pi \hbar }{ml}(n -  \frac{\Phi }{\Phi _{0}}) \eqno{(13b)}$$
in the long segment according to (8) and (9). According to (13a) the critical current of the ring with the strong asymmetry should be equal 
$$I_{c} = |\pm I_{c0} +  \frac{l}{l_{long}}I_{p}| = |\pm I_{c0} +  \frac{l}{l_{long}}I_{p,A}2(n -  \frac{\Phi }{\Phi _{0}})| \eqno{(14a)}$$ 
at any magnetic flux value $|n - \Phi /\Phi _{0}| \leq 0.5$. Whereas the critical current of the ring with a weak asymmetry $(l_{u} - l_{l})l_{u}/l^{2} \geq  I_{p,A}/I_{c0}$ should be equal
$$I_{c} =  |\pm \frac{l_{long}}{l_{sh}}I_{c0} -  \frac{l}{l_{sh}}I_{p}| \eqno{(14b)}$$
in some region near $\Phi  = (n'+0.5)\Phi _{0}$, where the pair velocity in the long segment exceeds the one in the short segment $|v_{u}| = v_{c} > |v_{l}|$ due to the persistent current. The jump $\Delta  I_{c} = \Delta  I_{p}l/l_{long}$ of the critical current of the ring with the strong asymmetry should exceed the jump of the persistent current according to (14a). 

But the jumps are not observed at the measurement of aluminium ring with the strong asymmetry, Fig.3. The jump is absent as in the case of an asymmetric ring, but for a different reason. The experimental dependence $I_{c+}(\Phi /\Phi_{0})$ and $I_{c-}(\Phi /\Phi_{0})$ are not shifted and correspond to the theoretical predictions (14a) near integer values of the flux quantum $|n - \Phi /\Phi _{0}| < 0.25$, Fig.3. The jump is absent due to the deviation of the measured values of the critical current from the theoretically predicted values near the half of the flux quantum $|n - \Phi /\Phi _{0}| > 0.25$, Fig.3. Therefore, the smooth change in the critical current $I_{c+}(\Phi /\Phi_{0})$, $I_{c-}(\Phi /\Phi_{0})$ is observed instead of the jump that should be observed when changing the quantum number by the unit from $n = n'$ to $n = n'? + 1$ at $\Phi  = (n'?+0.5)\Phi _{0}$. We can't doubt that the quantum number $n $ changes because the measured values at $|n - \Phi /\Phi _{0}| < 0.25$ are described by a theoretical dependence (14a) in which $n = -2$ at $\Phi /\Phi_{0} = -2$; $n = -1$ at $\Phi /\Phi_{0} = -1$; $n = 0$ at $\Phi /\Phi_{0} = 0$; $n = 1$ at $\Phi /\Phi_{0} = 1$; $n = 2$ at $\Phi /\Phi_{0} = 2$, Fig.3. 

The results of the measurements of the critical current of the superconducting ring with asymmetric link-up of current leads $l_{u} > l_{l}$ at $|n - \Phi /\Phi _{0}| > 0.25$, Fig.3, are very strange since the quantum number $n$ must be an integer number according to the very foundation of the quantum theory - the Bohr quantisation (1). The $I_{c+}(\Phi /\Phi_{0})$ values measured at $|n' - \Phi /\Phi _{0}| > 0.25$, Fig.3, corresponds to the values of the velocity of Cooper pairs which must be forbidden according to the quantization condition (5). The equality of the $I_{c+}(\Phi /\Phi_{0})$ values for the magnetic flux $\Phi  = n'\Phi _{0}$ and $\Phi  \approx  (n'+0.5)\Phi _{0}$, Fig.3,  means that the circular velocity (5) must be equal to zero for both $\Phi  = n'\Phi _{0}$ and $\Phi  \approx (n'+0.5)\Phi _{0}$. The state $n = n'$ with the zero velocity is permitted at $\Phi  = n'?\Phi _{0}$ but such state is forbidden at $\Phi  \approx (n'?+0.5)\Phi _{0}$ according to the quantization condition (5). The velocity (5) can be equal zero at $\Phi  = (n'+0.5)\Phi _{0}$ only if the quantum number can be non-integer $n = n' + 0.5$. Therefore the absence of the jump in the results of measurements of the critical current of the superconducting ring with asymmetric link-up of current leads $l_{u} > l_{l}$ may be an important experimental result having fundamental importance. 

\begin{figure}
\includegraphics{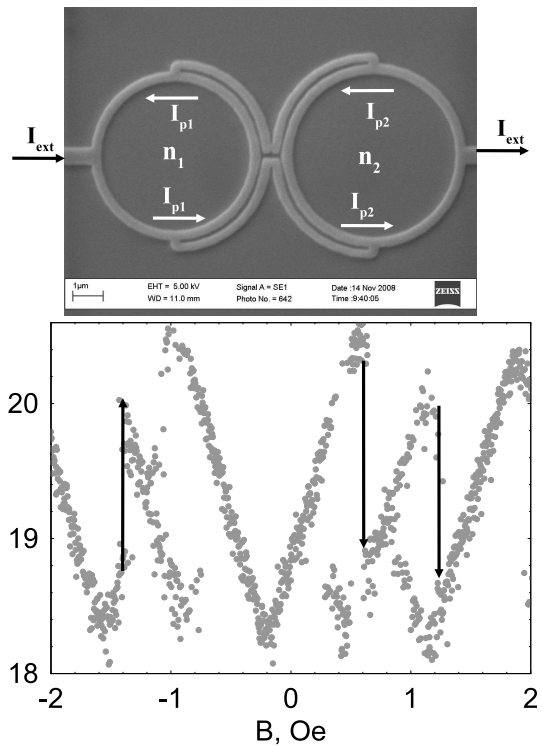}
\caption{\label{fig:epsart} The upper picture: Photo of the aluminum structure on which the jumps of the critical current were observed. The diameters of the rings are $r_{1} \approx 2.1 \ \mu m$ and $r_{2} \approx 2.3 \ \mu m$. The area of the ring $S_{1} = \pi r_{1}^{2} \approx 13.9  \ \mu m^{2}$ and $S_{2} = \pi r_{2}^{2} \approx 16.6  \ \mu m^{2}$ correspond to the different periods of the oscillations in magnetic field $B_{0,1} = \Phi_{0}/S_{1} \approx  1.5 \ Oe $ and $B_{0,2} = \Phi_{0}/S_{2} \approx  1.2 \ Oe $. The width of the rings is $ \approx 0.4 \ \mu m$, the width of the lines connecting the rings is $ \approx 0.3 \ \mu m$, and the film thickness is $d \approx 30 \ nm$. The lower picture: The results of measurement of the critical current of the structure shown in the upper figure. The jumps of the critical current because of the change of the quantum number of one of the rings are indicated with the arrows.}
\end{figure}

\section{The jumps associated with changes in the quantum number}
\label{}
The strangeness of the absence of the critical current jumps in measurements of superconducting rings with asymmetry $s_{u} > s_{l}$ and $l_{u} > l_{l}$ is enhanced by the observation of the jumps associated with changes in the quantum number in other cases. The jumps are observed in the results of the measurements of the magnetic flux  $\Delta \Phi _{Ip} = L_{f}I_{p}$ induced by the persistent current $I_{p}$ of the flux qubit superconducting loop interrupted by three Josephson junctions \cite{Tanaka2002}. Although the observations of a $\chi $-shaped crossing of the $I_{p,n}( \Phi)$ and $I_{p,n+1}( \Phi)$ dependencies, see Fig.3 in \cite{Tanaka2009} contradicts to the theoretical prediction of quantum mechanics. The jumps of the critical current and the voltage connected with the change of the quantum numbers of two superconducting loops were observed in \cite{NanoLet2017} at measurements of the superconducting differential double contour interferometer - two identical superconducting loops connected with the help of two Josephson junctions in two points.

The structure, shown in Fig.4, is similar to the superconducting differential double contour interferometer (DDCI), investigated in \cite{NanoLet2017}, in the sense that the wave functions describing the superconducting state in its two rings are connected at two points. The rings of this structure are connected by superconducting strips, not by Josephson transitions as in the case of the DDCI \cite{NanoLet2017} and the two rings are located next to each other, not on top of each other as in the case of of the DDCI \cite{NanoLet2017}. Therefore this structure can be considered as intermediate between single superconducting rings in which the jumps associated with the changes in the quantum number are not observed and the DDCI in which such jumps are definitely observed \cite{NanoLet2017}. The measurements have revealed that the jumps of the critical current of this structure are observed, Fig.4. However, they cannot be linked to the changes in the quantum numbers $n_{1}$ and $n_{2}$ of the two rings with the same reliability as is done in the case of the DDCI \cite{NanoLet2017}. 

\section{Conclusion}
Nanostructures have the advantage over atoms that they can be made of various sizes and diversified shapes. This opens up great opportunities for the experimental investigations of the problem of the quantum jumps. The results of the first experiments, presented in this work, testify that such investigations may have fundamental importance. The paradoxical absence of the quantum jumps of the critical current of the superconducting asymmetric rings $s_{u} > s_{l}$ and the rings with asymmetric link-up of current leads $l_{u} > l_{l}$ poses the task of a theoretical explanation of this paradox. The question, why jumps are not observed despite the change in the quantum number, which is obvious from the observed quantum periodicity, remains unanswered. For a more reasonable answer to this question, additional experimental research is needed.  The measurements of the rings of different superconductors with various sizes and diversified shapes may clarify the question of the universality of the revealed absence of the quantum jumps of the critical current. The results of such measurements may be of fundamental importance for a better understanding of the fundamentals of quantum mechanics and in particular macroscopic quantum phenomena.   

\section*{Acknowledgement} This work was made in the framework of State Task No 075-00920-20-00.

\end{document}